\documentclass[11pt]{article}

\usepackage{amsmath}
\usepackage{amsfonts}
\usepackage{amssymb}
\usepackage{latexsym}
\usepackage{cite}
\usepackage{lscape} 

\renewcommand{\d}{\mathrm{d}}

\DeclareMathSymbol{\mg}{\mathrel}{symbols}{"1D}

\newcommand{\ga}{\alpha}
\newcommand{\gb}{\beta}
\renewcommand{\gg}{\gamma}
\newcommand{\gd}{\delta}
\renewcommand{\ge}{\epsilon}

\newcommand{\gf}{\phi}

\newcommand{\gm}{\mu}
\newcommand{\gn}{\nu}

\newcommand{\gk}{\kappa}
\newcommand{\gl}{\lambda}
\newcommand{\gr}{\rho}

\newcommand{\gs}{\sigma}

\newcommand{\go}{\omega}

\newcommand{\gp}{\pi}
\newcommand{\gps}{\psi}
\newcommand{\get}{\eta}

\newcommand{\gG}{\Gamma}

\newcommand{\gL}{\Lambda}

\newcommand{\gO}{\Omega}

\newcommand{\cD}{{\cal D}}

\newcommand{\tm}{{\tilde m}}

\newcommand{\tT}{{\tilde T}}

\newcommand{\tr}{\text{tr}}

\newcommand{\lra}{\leftrightarrow}

\newcommand{\ra}{\rightarrow}

\newcommand{\der}{\partial}

\newcommand{\inv}{^{-1}}
\newcommand{\dsp}{\displaystyle}

\newcommand{\labl}[1]{\label{#1}}
\newcommand{\beq}{\begin{equation}}
\newcommand{\eeq}{\end{equation}}
\newcommand{\barr}{\begin{array}}
\newcommand{\earr}{\end{array}}
\newcommand{\equ}[1]{\begin{gather} #1 \end{gather}}

\newcommand{\arry}[2]{\begin{array}{#1} #2 \end{array}}

\newcommand{\pmtrx}[1]{\begin{pmatrix} #1 \end{pmatrix}}

\newcounter{oldcounter}

\newcommand{\bgps}{{\bar\psi}}

\newcommand{\tgf}{{\tilde \phi}}

\newcommand{\ba}[2]{\[\begin{array}{#2}\label{#1}}
\newcommand{\ea}{\end{array}\]}
\newcommand{\be}{\begin{equation}}
\newcommand{\ee}{\end{equation}}
\newcommand{\bea}{\begin{eqnarray}}
\newcommand{\eea}{\end{eqnarray}}

\newcommand{\U}[1]{\mathrm{U(#1)}}
\newcommand{\SU}[1]{\mathrm{SU(#1)}}
\newcommand{\SO}[1]{\mathrm{SO(#1)}}

\begin{document}

\thispagestyle{empty}

\begin{flushright}
FTPI-MINN-04/43 \\ 
UMN-TH-2329/04  \\      
hep-th/0411184
\end{flushright}
\vskip 2 cm
\begin{center}
{\Large {\bf 
Chiral gravity as a covariant formulation of massive gravity 
}
}
\\[0pt]
\vspace{1.23cm}
{\large
{{\bf Stefan Groot Nibbelink$^{a,}$\footnote{
{{ {\ {\ {\ E-mail: nibbelin@hep.umn.edu}}}}}}}, 
{\bf Marco Peloso$^{b,}$\footnote{
{{ {\ {\ {\ E-mail: peloso@physics.umn.edu}}}}}}},
\bigskip }\\[0pt]
\vspace{0.23cm}
${}^a$ {\it 
William I. Fine Theoretical Physics Institute,}
${}^b$ {\it 
School of Physics \& Astronomy, \\
University of Minnesota, 116 Church Street S.E., 
Minneapolis, MN 55455, USA 
\\}
}
\bigskip
\vspace{1.4cm} 
\end{center}
\subsection*{\centering Abstract}

We present a covariant nonlinear completion of the Fierz--Pauli (FP)
mass term for the graviton. The starting observation is that the FP
mass is immediately obtained by expanding the cosmological constant
term, i.e. the determinant of the vielbein, around Minkowski space to
second order in the vielbein perturbations. Since this is an unstable
expansion in the standard case, we consider an extended
theory of gravity which describes two vielbeins that give rise to chiral
spin--connections (consequently, fermions of a definite chirality only couple
to one of the gravitational sectors). As for Einstein gravity with a
cosmological constant, a single fine--tuning is needed to recover a
Minkowski background; the two sectors then differ only by a constant
conformal factor. The spectrum of this theory consists of a massless
and a massive graviton, with FP mass term. The theory possesses
interesting limits in which only the massive graviton is coupled to
matter at the linearized level.

\newpage 

\setcounter{page}{1}

\section{Introduction}
\labl{sc:intro}

%
%
There is compelling evidence that the expansion of the Universe is
presently accelerating~\cite{acc}. The simplest cause for this
effect, a vacuum energy $\Lambda \sim 10^{-120} \, M_P^4 \,$, is
so unnaturally small that it is worth searching for alternative
explanations. The most straightforward possibility is that, due to some
(unknown) symmetry, $\Lambda = 0 \,$, and that the present
acceleration is instead due to some (quintessence) field whose equation of state is
sufficiently close to the one of vacuum, $w \leq - 0.7\,$~\cite{acc}. 
An alternative, more drastic, possibility, is that the
present accelerated expansion signals a modification of standard
gravity at very large scales. A well--studied modification is massive
gravity, which is expected to be weaker at distances larger than the inverse
graviton mass $m_g^{-1}\,$ by Yukawa suppression. If the graviton mass
is comparable with the present Hubble parameter $H_0 \,$, this may
lead to accelerated expansion at the largest observable
scales. A different road to modify the gravitational interactions is
to introduce a ``nonstandard'' form of matter, which, once coupled to a
conventional gravitational sector, can change the properties of the
graviton. One recent example is the theory of
ghost condensation~\cite{Arkani-Hamed:2003uy},
where the Lorentz symmetry is broken by the gradient of a scalar
field.~\footnote{The phenomenology of this model in the physically
relevant cases has been studied in~\cite{dps}.} Although these
proposals are quite speculative, they are certainly
suggestive directions worth further investigations. 
The quest for consistent modifications of standard gravity is by itself a
very interesting and nontrivial theoretical subject, which has indeed
drawn considerable attention both in the past and at present.

%
%
The structure of the mass term for the graviton was
investigated by Fierz and Pauli~\cite{fipa} already in 1939.
They showed that the only Lorentz--invariant ghost--free
mass term for a graviton in
Minkowski background~\footnote{As discussed in~\cite{rudu}, a richer
structure of ghost--free mass terms is possible if one is willing to
give up Lorentz invariance.} is  
\begin{equation}
\frac 12 \, m_g^2 \, \eta^{\mu \nu} \, \eta^{\alpha \beta} \Big[ h_{\mu \nu} \,
h_{\alpha \beta} - h_{\mu \alpha} \, h_{\nu \beta} \Big]~,  
\label{fp}
\end{equation}
where $h_{\mu \nu} = g_{\mu \nu} - \get_{\mu \nu}$ defines the metric
perturbations. Because this Fierz--Pauli (FP) term~(\ref{fp}) is not
covariant, the graviton acquires three additional degrees of freedom,
that are instead gauge modes for standard gravity. The divergence--free vector
components do not couple to conserved sources in the massless limit. 
However, the third (longitudinal) component is sourced by the trace of
the energy--momentum tensor, and it does not decouple in this limit. 
Consequently, the graviton propagator exhibits a discontinuity at the
linearized level between the massless and the massive case~\cite{vdvz}. 
The main phenomenological consequence of this discontinuity would
be a too large modification of the bending of the light from the sun.
However, nonlinear effects drastically change this picture:  As
realized in~\cite{arkady1} (and re--examined in~\cite{arkady2,ags}), 
gravity mediated by a massive graviton becomes strong at macroscopic
distances.  By extending the Schwarzschild  solution to the FP case,
one can show that, while the linearized (one graviton exchange)
approximation is indeed discontinuous, the full nonperturbative
solution has a smooth $m_g \rightarrow 0$ limit~\cite{arkady1}, {at
least at some finite range of distances from the source. It is however
unclear whether the solution found in~\cite{arkady1}
can be extended into a regular solution from the source up to infinity; 
some specific nonlinear completions of~(\ref{fp}) have been analyzed
-- partially through numerical calculations -- in~\cite{damour}, and
the solutions were found to develop singularities at finite distance
from the source.  However, it is hard to see whether this conclusion
is true for arbitrary completions. 

Therefore, the FP mass term~(\ref{fp}) can not be ruled out by the
discontinuity present at the linear analysis. Unfortunately, going
beyond the linear regime one faces the problem of the strong
sensitivity to the completion of massive gravity. This is
particularly true if we think to massive gravity as an effective field
theory, where all higher order interactions not forbidden by any
symmetry should be included. Among the symmetries restricting these
higher order terms,  one can easily argue that the nonlinear
completion should be covariant. As shown in~\cite{bode} (see
also~\cite{gagr}) FP gravity has a clear instability at the nonlinear
level, related to the fact that, with the FP term present, the
Hamiltonian of the system is not positive semi--definite. This problem
seems to persist for generic nonlinear completions of~(\ref{fp}). In a
covariant theory, 
like the Einstein gravity, the Hamiltonian defines a constraint, so
that this instability is not present from the outset. Hence, one may
expect that a covariant non--linear completion of the FP mass term
does not suffer from this specific nonlinear instability.}

%
%
When one is considering modifications of gravity, one may contemplate
other possibilities beside a possible graviton mass. One rather exotic
question is whether a theory of gravity can make a distinction between
fermions of different chirality. This is not an unnatural question in light of
the Standard Model (SM) of particle physics, since the Electroweak
interaction couples differently to left-- and right--handed quarks and
leptons. This is technically implemented using 
a chiral gauge connection and exploiting that some fermions are
$\SU{2}_L$ doublets while others are singlet. Applying such arguments
to the standard theory of gravity shows that this is not possible in
an interesting way: Gravity couples to the spin of fermions via the
spin--connection contracted with the spin generators $\gamma_{ab}$ of
the local Lorentz group (since fermions do not transform under general
coordinate transformations, but under local Lorentz transformations). 
Of course, one could make the interaction chiral by contracting the
spin--connection with the spin generator of a given chirality (either
$\frac{1+\gamma_5}{2}$ or $\frac{1-\gamma_5}{2}\,$), but
then fermions of the opposite chirality would not interact with the
spin--connection at all! Therefore, to obtain an interesting theory of
chiral gravity one needs two independent spin--connections. This in
turn implies that one needs a theory with two dynamically independent
vielbeins $e_{\pm\,\gm}{}^a$ out of which the spin--connections of the
$\pm$ chiralities can be constructed. We conclude that a theory of
chiral gravity~\footnote{Chiral gravity should not be confused with
spinor gravity \cite{hw} in which gravity arises from fermion
dynamics.} is necessarily a theory of bi--gravity. 

%
%
To see how these two seemingly unrelated modifications of conventional
gravity come together, we review why it is hard to obtain a covariant
completion of the FP mass term. The 
main reason is that there is only one nontrivial scalar quantity which
can be constructed from a metric without using derivatives, namely its
determinant. Suppose we start from the simplest possibility,
\begin{equation}
S = \int d^4 x \sqrt{- g} \, \left[ \frac{1}{2} \, M_P^2 \, R -
\Lambda \right]~. 
\label{azds}
\end{equation}
Expanding the cosmological constant term up to second
order in the perturbations $h_{\mu \nu}$ around Minkowski background
gives: 
\begin{eqnarray}
\sqrt{- {\rm det} \left( \eta + h \right)} &=& 1 + \frac{1}{2} \,
\eta^{\mu \nu} \, h_{\mu \nu}
+ \frac{1}{4} \, \eta^{\mu \nu} \, \eta^{\alpha \beta} \Big[
\frac{1}{2} \, h_{\mu \nu} \, h_{\alpha \beta} - h_{\mu \alpha} \,
h_{\nu \beta} \Big] + {\cal O}( h^3)~. 
\label{expdet}
\end{eqnarray}
Immediately two problems appear when we compare this
expansion with the FP term: 
(a) The quadratic terms are not those of the FP mass term.  
(b) The linear term signals that we are expanding around a wrong
background (Minkowski rather than de Sitter, as we should be doing 
for~(\ref{azds})). 
These observations (and their extension to an arbitrary function
of the determinant) led~\cite{bode} to conclude that completions
of~(\ref{fp}) cannot be covariant: A background metric
$g^{(0)}{}_{\gm\gn}$ has to be combined with the metric $g_{\gm\gn}$ in
order to form nontrivial expressions which can reduce to~(\ref{fp}) in the
weak field limit.~\footnote{One way to evade the
arguments of~\cite{bode}  is to consider gravity
in extra dimensions, and to preserve covariance for the standard $3+1$
coordinates. Conventional KK theories have a tower of massive gravitons;
however the massless graviton is also present, so that one does not recover
a theory of a single massive graviton~\cite{Duff}. Interesting progresses
have been recently made in braneworld models.
In particular, the model~\cite{dgp} (see also~\cite{grs}
for an interesting earlier attempt) shares many analogies with $4$
dimensional models of massive gravity. }
We will now show how the chiral theory of gravity
introduced above can help resolve these problems.

%
%
Let us first address the problem of the
appearance of the non--FP quadratic term in~(\ref{expdet}). 
Our starting observation is that the FP coefficients are obtained if one
considers first order perturbations $F_{\mu \nu}$ in the vielbein 
$e_\mu{}^a = \delta_\mu^a \, + F_{\mu \nu} \, \eta^{\nu a}\,$
(rather than first order perturbations $h_{\mu \nu}$ of the metric): 
\begin{equation}
e = \sqrt{-g} = 
1 + \eta^{\mu \nu} F_{\mu \nu} + \frac{1}{2} \, \eta^{\mu \nu}
\, \eta^{\alpha \beta} \Big[ F_{\mu \nu} \, F_{\alpha \beta} -
F_{\mu \alpha} \, F_{\nu \beta} \Big] + {\cal O}(F^3)~.
\label{expdet2}
\end{equation}
The reason is an extra contribution from the linear term in~(\ref{expdet}), 
due to the second order perturbations in the metric 
\begin{equation}
g_{\mu \nu} = e_\mu{}^a \, \eta_{a b} \, e_\nu{}^b = 
\eta_{\mu \nu} + 2 \, F_{\mu \nu} + \eta^{\lambda
\sigma} \, F_{\mu \lambda} \, F_{\nu \sigma}~. 
\label{quadh}
\end{equation}
Clearly, one could have started directly from the expansion~(\ref{quadh}),
 but this would have been a  very special and unmotivated choice. We
find it remarkable that the FP mass term simply follows from the
natural expansion~(\ref{expdet2}) of the determinant of the vielbein.

%
%
The remaining issue is how to eliminate the linear term in the
expansion~(\ref{expdet2}) when one expands around a Minkowski 
background. We will show that this can be done by the bi--gravity
theory that corresponds to chiral gravity, containing two vielbeins
$e_{\pm\,\gm}{}^a\,$ and metrics $g_{\pm\,\gm\gn}$. This theory
consists of a copy of the standard 
action~(\ref{azds}) for each sector with their own Planck mass and
cosmological constant. In addition, there is an interaction that couples
the two vielbeins in a covariant way via a term which, though similar to,
cannot be represented as a determinant. The Minkowski background is
not introduced by hand as in~\cite{bode}, but it arises dynamically by
a subtle balance between the two gravitational sectors, given by a 
single fine--tuning of cosmological constant--like parameters. 
This constitutes the same amount of fine--tuning as for the
cosmological constant in the standard theory of gravity.

%
%
Combining these ingredients we obtain a theory, that at the linearized
level describes two graviton modes which are coupled via a FP mass
term.  The FP mass can have generic value since the sizes of the
cosmological constants are not fixed. 
The spectrum of the linearized perturbations consists of a
massless graviton and a massive graviton, with a FP mass term. 
(The presence of a single massless graviton is to be expected because
the theory of chiral gravity is covariant.)
We treat matter in the simplest way possible, that is by coupling
it either to the metric $g_{+\,\gm\gn}$ or to $g_{-\,\gm\gn}\,$; this defines 
two matter sectors. Since the massive/massless gravitons are linear
combinations of the perturbations of the two vielbeins, these mass
eigenstates couple to both matter sectors. The interactions of the
massless graviton with the two sectors turn out to be equal. 
The ones of the massive graviton are instead of different strengths and
of opposite sign (mediating a gravitational repulsion between the two
sectors). The theory possesses interesting
limits in which the massless graviton decouples at the linear order,
and the massive graviton only couples to one of the two matter
sectors. Hence, if we include all the matter in this sector, this
limit realizes a covariant nonlinear completion of the FP mass
term~(\ref{fp}). 

%
%
The plan of this paper is the following. Section~\ref{sc:vielbein}
reviews the formulation of standard Einstein gravity in terms of the
vielbein and the spin--connection. For this we use a presentation
employing Clifford algebra valued differential forms, since it
naturally generalizes to the theory of chiral gravity described in
Section~\ref{sc:chigrav}. Readers who are primarily interested in the
background evolution and the linearized spectrum can skip these more
technical sections, since the discussion of the final action of chiral
gravity in Section~\ref{sc:background} is self--contained. 
The main purpose of this section is to derive and describe
corresponding de Sitter and Minkowski background solutions.  
The spectrum of perturbations around the
Minkowski background is computed in Section~\ref{sc:perturs}. 
The coupling of chiral gravity to matter is discussed  in
Section~\ref{sc:matter}. In particular we pay attention to how the SM
can be coupled to this theory. From the coupling of the graviton mass
eigenstates to matter we derive effective Planck masses, and limits in
which the massless graviton decouples. Section~\ref{sc:concl} presents our
conclusions and some outlook for future work. Appendix
\ref{sc:details} contains our notation and a few technical results.

\section{Gravity using the vielbein formalism}
\labl{sc:vielbein}

We review an elegant representation of the theory of gravity using the
vielbein formalism and differential forms. We will use the same
formalism to introduce our proposal for a theory of chiral gravity in
the next Section. Our conventions and notations have been collected in
appendix~\ref{sc:details}. We have followed the presentation of
the vielbein formalism of~\cite{vanh}, and we have used Clifford
algebra valued forms similar to \cite{cham}. 
Our starting point is Einstein gravity with a
cosmological constant $\gL$, described by vielbein and
spin--connection one--forms
\equ{
e_1 = e_\gm{}^a\, \gg_a\, \d x^\gm~, 
\qquad 
\go_1 = \frac 14 \go_\gm^{ab}\, \gg_{ab}\, \d x^\gm~,
\labl{VielBSpinC}
} 
which are taken to be independent. The action can be written as 
\equ{
S = \frac i4 \int \tr\, \gg_5 \Big( 
M_P^2 \, e_1^2 \, R_2(\go) - \frac 1{4!} \gL\, e_1^4
\Big)~,
\labl{VielGrav}
} 
with the Planck mass $M_P^2 = 1/(8 \gp G)\,$, and the curvature
associated to the spin--connection 
\equ{
R_2(\go) = \d \go_1 + \go_1^2
= \frac 18 R^{ab}_{\gm\gn}(\go) \, \gg_{ab}\, \d x^\gm \d x^\gn~. 
\labl{SpinCurv}
}
The equation of motion of the spin--connection is fulfilled, if it
satisfies the Maurer--Cartan equation 
\equ{
\d e_1 + e_1 \go_1 + \go_1 e_1 = 0~. 
\labl{MaurerCartan} 
}
Writing this equation out in components implies that there is a
torsion--free connection, $\gG^\gl_{\gm\gn} = \gG^\gl_{\gn\gm}\,$, such
that the covariant derivative $\cD_\gm$ is covariantly constant on the
vielbein: 
\equ{
\cD_\gm e_\gn{}^a = \der_\gm e_\gn{}^a - \gG^\gl_{\gm\gn} e_\gl{}^a 
+ \go^{ab}_\gm \get_{bc} e_\gn{}^c = 0~. 
\labl{CovConst}
} 
We denote the solution of the spin--connection by 
$\go_\gm^{ab} = \go^{ab}_\gm(e)\,$, its explicit form is given in 
\eqref{solSpinCon}. Substituting this expression back into
\eqref{CovConst} one finds that $\gG^\gl_{\gm\gn}$ is precisely the
conventional Christoffel--connection $\gG^\gl_{\gm\gn}(g)$ defined
from the metric \eqref{quadh}. 
The curvature $R^\gl{}_{\gr\gm\gn}(\gG)$ associated to the connection 
(which is given explicitly in \eqref{ChrCon}) 
can be expressed in terms of the curvature \eqref{SpinCurv} of the
spin--connection as 
\equ{
R^\gl{}_{\gr\gm\gn}(\gG) = 
R^{ab}_{\gm\gn}(\go)\, (e\inv)_a{}^\gl\, \get_{bc}\, e_\gm{}^c~,  
\labl{EqCurvs}
}
by working out the commutator $[\cD_\gm, \cD_\gn] e_\gr{}^a = 0$ in
components. Therefore, we may refer to the curvature, Ricci--tensor 
$R^a_\ga = R^{ab}_{\ga\gb} (e\inv)_b{}^\gb$ and
the curvature scalar without indicating whether the spin-- or
Christoffel--connection is used. By employing some Clifford and form
algebra, one can show that \eqref{VielGrav} turns into the standard
action for gravity with a cosmological constant \eqref{azds}.

Before we move on to describe our chiral theory of gravity, we would
like to make a few comments concerning the symmetries of the standard
theory of Einstein gravity in the vielbein formalism: Since the action
\eqref{VielGrav} was written in terms of differential forms only, it
is invariant by construction under general coordinate
transformations. In addition, the theory is invariant under local
$\SO{1,3}$ Lorentz transformations
\equ{
e_1 \ra \gO \, e_1\, \gO\inv, 
\qquad 
\go_1 \ra \gO\, \big( \go_1 + \d \big)\, \gO\inv~, 
\qquad 
R_2(\go) \ra \gO\, R_2(\go)\,  \gO\inv~, 
\labl{locLor}
}
with $\gO = \exp \frac 14 \gO^{ab}\, \gg_{ab}\,$. Finally, we note
that the vielbeins are only defined up to a sign, $e_\gm{}^a \ra -
e_\gm{}^a$ being a symmetry of the theory.

\section{Chiral gravity}
\labl{sc:chigrav}

After the review of the standard theory of gravity in the vielbein and
differential form representation, we describe our proposal for a
chiral theory of gravity. The Electroweak sector of Standard Model of
particle physics is a chiral theory, i.e.\ the gauge fields couple
differently to left-- and right--handed states. In the formalism, this
is realized by having chiral $\SU{2}_L \times \U{1}_Y$ gauge
connections. By analogy, we introduce two vielbeins $e_{\pm\,
\gm}{}^a$ and two spin--connections $\go_{\pm\, \gm}^{ab}\,$. Since
only the spin--connection one--form \eqref{VielBSpinC} preserves
chirality, as it is proportional to $\gg_{ab} \,$, we take only the
spin--connections to be chiral:
\equ{
e_{1\pm} = e_{\pm\, \gm}{}^a\, \gg_a\, \d x^\gm~,
\qquad 
\go_{1\pm} = \frac 14 \go_{\pm\,}{}_\gm^{ab}\, \gg_{ab}
\, \frac {1\pm \gg_5}2\, \d x^\gm~.
\label{eochi}
}
We require that these vielbeins and spin--connections transform in the
appropriate way under local Lorentz transformations
\equ{
e_{1\pm} \ra \gO \, e_{1\pm} \, \gO\inv~, 
\qquad 
\go_{1\pm} \ra \gO\, 
\Big( \go_{1\pm}  +  \frac {1\pm \gg_5}2 \,\d \Big)\, 
\gO\inv~.
}
In principle, we could have required that both $\pm$--sectors have
independent Lorentz transformations $\gO_\pm\,$. However, this
$\SO{1,3}_+ \times \SO{1,3}_-$ gauge symmetry would forbid any
interaction between the two sectors, and hence it would not lead to
interesting physics. Finally, we generalize the reflection symmetry of
the vielbein to both vielbeins $e_{+\,\gm}{}^a$ and $e_{-\,\gm}{}^a$
independently.

In view of the field content, the natural generalization of
\eqref{VielGrav} is given by 
\equ{
S = \frac i4 \int \tr  \gg_5 \Big\{
M_+^2 \, e_{1+}^2 \, R_{2+}(\go_+) 
+ M_-^2 \, e_{1-}^2 \, R_{2-}(\go_-) 
- \frac 1{4!} \Big( 
\gL_+\, e_{1+}^4  -2 \gL_0\, e_{1+}^2 e_{1-}^2  + \gL_-\, e_{1-}^4
\Big) 
\Big\}~,
\labl{ChiVielGrav}
}
where $R_{2\pm}(\go_\pm) = \d \go_{1\pm} + \go_{1\pm}^2\,$. Here
$M_\pm$ can be thought as the analogies of the Planck masses for the
$\pm$--sectors of gravity, and $\gL_\pm$ and $\gL_0$ parameterize all
possible cosmological constants compatible with local Lorentz
invariance and the vielbein reflection symmetries. The sign
conventions for the cosmological constants will become clear in the
next Section, where we investigate the background solutions. Moreover,
we will assume throughout this work that $\gL_0\neq 0\,$, i.e.\ that
the $\SO{1,3}_+ \times \SO{1,3}_-$ Lorentz symmetry is explicitly
broken to its diagonal subgroup, which can be identified with the
Lorentz group of conventional gravity.

The kinetic terms do not 
represent the most general form compatible with our symmetries. We
restrict ourselves to these structures only, since for them one can
immediately give expressions for the spin--connections in terms of the
vielbeins, using the solution $\go_\gm^{ab} = \go_\gm^{ab}(e)$ encoded
in the Maurer--Cartan equation \eqref{MaurerCartan} for standard
gravity. (The explicit solution is given in \eqref{solSpinCon}.) In
particular, the spin--connection one--forms and the curvature
two--forms are given by  
\equ{
\go_{1\pm} = \frac 14 \go_\gm^{ab}(e_\pm) 
\gg_{ab} \frac {1\pm\gg_5}2 \d x^\gm~, 
\qquad 
R_{2\pm}(\go_\pm) = \frac 18 R_{\gm\gn}^{ab}\big(\go(e_\pm)\big) 
\gg_{ab} \frac {1\pm\gg_5}2 \d x^\gm \d x^\gn~. 
}
Also in direct generalization of the situation in standard gravity,
\eqref{quadh}, one can introduce the metrics 
\equ{
g_{\pm\,\gm\gn} = e_{\pm\, \gm}{}^a\,\get_{ab}\,e_{\pm\,\gn}{}^b~,
\label{metrics}
}
the Christoffel--connections 
$\gG_{\pm}{}_{\gm\gn}^\gl = \gG^\gl_{\gm\gn}(g_\pm)$ and 
curvatures $R^\gl_{\gr\gm\gn}(\gG_\pm)\,$. In particular, we find that
the identification between the curvatures in terms of the Christoffel--
and spin--connections \eqref{EqCurvs} holds for both sectors
separately. This has an important consequence: The first two terms in
\eqref{ChiVielGrav} give rise, aside from the conventional kinetic
terms, to a part without $\gg_5$: 
\equ{
\pm \frac i4 M_\pm^2 \int\tr\, e_{1\pm}^2 R_2(\go_\pm) 
= \mp 2i \int \d^4 x\, \ge^{\ga\gb\gg\gd} R_{\ga\gb\gg\gd}(g_\pm)~,
}
which would make the theory non--unitary. However, this term vanishes
because of the cyclicity of the Riemann tensor $R_{\ga\gb\gg\gd}(g)$.

To evaluate the cosmological constant--like terms we set up some
additional notation. Let $A, \ldots ,D$ be four matrices, which like
the vielbeins, carry one spacetime and one tangent space 
index. We define 
\equ{
\langle A B C D \rangle = -\frac 1{4!} 
\ge^{\ga\gb\gg\gd} \ge_{abcd} A_\ga^a B_\gb^b C_\gg^c D_\gd^d~. 
\label{newproduct}
}
The ordering of the matrices $A,\ldots,D$ is irrelevant in this
expression, and it generalizes the notion of a determinant, in the
sense that $\langle A^4 \rangle = \det(A)\,$. However, 
$\langle A^2 B^2\rangle$ cannot be written as a determinant. 
The cosmological constant terms in \eqref{ChiVielGrav} can
be cast in the form  
\equ{
\frac i4 \frac 1{4!} \int \tr\, \gg_5 e_{1+}^p e_{1-}^{4-p} 
= \int \d^4 x\, \langle e_+^p e_-^{4-p} \rangle~,
}
for $p=0, \ldots 4$. Similarly to \eqref{newproduct}, we define 
\equ{
\langle A B C  \rangle^\gd_d = -\frac 1{3!} 
\ge^{\ga\gb\gg\gd} \ge_{abcd} A_\ga^a B_\gb^b C_\gg^c~. 
}
In this notation, the Einstein equations of chiral gravity can be
compactly written as 
\equ{
\frac 14 M_\pm^2 \, \ge^{\ga\gb\gg\gd} \ge_{abcd}\, 
e_{\pm\, \gb}{}^b R_\pm{}_{\gg\gd}^{cd} 
+ \gL_\pm \langle e_\pm^3 \rangle^\ga_a 
- \gL_0 \langle e_\pm e_\mp^2 \rangle^\ga_a = 0~. 
\labl{EqMot}
}
This representation of the Einstein equations will be our starting
point for our study of the perturbations in section \ref{sc:perturs}.

\section{Minkowski and de Sitter background solutions}
\labl{sc:background}

The action of the system can be rewritten in the form
\equ{
S = \int \d^4 x \left\{ \sqrt{- \, g_+} \left[ \frac{1}{2} M_+^2\, R_+
\, - \Lambda_+ \right] + 
 \sqrt{- \, g_-} \left[ \frac{1}{2} M_-^2\, R_- \, - \Lambda_- \right] 
+2 \, \Lambda_0 \, \langle e_+^2 \, e_-^2 \rangle \right\}~,
\label{acmetric}
}
where $g_+$ ($g_-$) is the determinant of the metric $g_{+ \,\mu \nu}$
($g_{- \,\mu \nu}$), and $R_+$ ($R_-$) its associated Ricci
scalar. The relations between the two metrics $g_{\pm\, \gm\gn}$ and
the respective vielbeins $e_\pm$ are given in~(\ref{metrics}).  The
first two terms describe two separate gravitational sectors,
characterized by the Planck masses $M_+$ and $M_-$, and the
cosmological constants $\Lambda_+$ and $\Lambda_-$. The two sectors
are coupled to each other via the last term (with $\langle \dots
\rangle$ defined in~(\ref{newproduct})). This coupling cannot be
written from the determinant of the two metrics, and this
theory is not equivalent to the bi--gravity theories usually considered
in the literature~\cite{kogan}.

We are interested in background solutions for (\ref{acmetric}) which
generalize the standard Minkowski and de Sitter solutions of Einstein
gravity. For this reason, we consider homogeneous and isotropic time
dependent vielbeins
\begin{equation}
e_{+ \, \mu}{}^a = {\rm diag } \big( a_+(t) ,\, b_+(t) ,\, b_+(t) ,\,
b_+(t) \big)~, \nonumber\\
\label{ansatz}
\end{equation}
and analogously for $e_{-\,\gm}{}^a$.  From this ansatz, one obtains
the following equations of motion
\begin{eqnarray}
&& \left( \frac{\dot{b}_+}{a_+} \right)^2 = \frac{1}{3 \, M_+^2} \,
\left( \Lambda_+ \, b_+^2 - \Lambda_0 \, b_-^2 \right)~, \label{e00p}
\\
&& \left( \frac{\dot{b}_+}{a_+} \right)^{\dsp \!\cdot} = \frac{1}{3 \, M_+^2}
\left( \Lambda_+ \, a_+ \, b_+ - \Lambda_0 \, a_- \, b_- \right)~,
\label{eiip} \\
&& \left( \frac{\dot{b}_-}{a_-} \right)^2 = \frac{1}{3 \, M_-^2} \,
\left( \Lambda_- \, b_-^2 - \Lambda_0 \, b_+^2 \right)~, \label{e00m}
\\
&& \left( \frac{\dot{b}_-}{a_-} \right)^{\dsp \!\cdot} = \frac{1}{3 \, M_-^2}
\left( \Lambda_- \, a_- \, b_- - \Lambda_0 \, a_+ \, b_+
\right)~. \label{eiim}
\end{eqnarray}
where dot denotes differentiation with respect to time. In addition,
the non--vanishing components of the spin--connection in the two
sectors (defined in eq.\ (\ref{eochi})) are given by 
\begin{equation}
\omega_{\pm \,}{}_{j}^{0 \, i} = \frac{\dot{b}_\pm}{a_\pm} \, \delta_j^i \,,
\end{equation}
where $i , j$ are spatial indices.

Eqs.~(\ref{e00p})-(\ref{eiim}) generalize the Friedman equations of
standard cosmology. For $\Lambda_0 = 0$ (two copies of the standard
case), the two equations~(\ref{eiip}) and~(\ref{eiim}) are actually
redundant, as a consequence of the Bianchi identities in the two
separate sectors (they can be replaced by the equations of state of
the fields driving the cosmological evolution, which in the present
case are simply $\Lambda_\pm =$ constant). Even in the presence of the
mix term, we can still exploit a generalized version of the Bianchi
identities, because the two Ricci scalars appearing
in~(\ref{acmetric}) are the standard ones. We obtain two simpler
equations than~(\ref{eiip})-(\ref{eiim}); by 
differentiating eq.~(\ref{e00p}) with respect to time, and by
combining it with eq.~(\ref{eiip}), we get
\begin{equation}
\frac{2 \, \Lambda_0 \, b_-}{a_+} 
\left( a_+ \, \dot{b}_- - a_- \, \dot{b}_+ \right) = 0~.
\end{equation}
An equivalent equation (wih $+$ and $-$ interchanged) is obtained in
the other sector. For $\Lambda_0 \neq 0 \,$ (we also assume that the
vielbeins are nonsingular) this enforces 
\begin{equation}
\frac{\dot{b}_+}{a_+} = \frac{\dot{b}_-}{a_-} 
\quad \Rightarrow \quad 
\omega_+ = \omega_-~.
\label{eqspin}
\end{equation}
Hence, the two background spin--connections are equal. This is a
very strong constraint, imposed by the rigid structure
of~(\ref{acmetric}) and by the simple background considered. 

To proceed, we equate the two right hand sides of 
eqs.\ (\ref{e00p})-(\ref{e00m}), and of eqs.\ (\ref{eiip})-(\ref{eiim}). 
Combining the two resulting algebraic equations gives 
\begin{equation}
a_+ \, b_- = a_- \, b_+~.
\label{relback}
\end{equation}
From the two eqs.~(\ref{eqspin}) and ~(\ref{relback}) we then find
\begin{equation}
\frac{b_+}{b_-} = \frac{a_+}{a_-} = C~,
\end{equation}
where $C$ is a (yet to be determined) constant. As for the
spin--connections, cf. eq.~(\ref{eqspin}), the symmetries of the
theory force the two sectors to expand with an equal rate. To
determine the background solutions explicitly, we specify a gauge for
the time variable (analogous to choosing physical or conformal time in
standard cosmology). Because 
$e_{1\pm} = a_\pm(t) \gg_0 \d t + \ldots\,$, time
parameterizations affect the product, but not the ratio between the
two ``lapse factors'' $a_\pm \,$. We fix this gauge freedom by setting
$a_+ \, a_- = 1 \,$, so that
\begin{equation}
a_\pm = C^{\pm \, 1/2}~.
\end{equation}
In this gauge, the time evolution is encoded by a single function 
$b\left( t \right)$, which we define as 
\begin{equation}
b_\pm (t) \equiv C^{\pm \, 1/2} \, b(t)~,
\qquad 
H \equiv \frac{\dot{b}}{b}~.
\end{equation}
The remaining equations of motion~(\ref{e00p}) and~(\ref{e00m})
determine the constant $C$ and the Hubble parameter: 
\begin{eqnarray}
C &=& \left( \frac{M_+^2 \, \Lambda_- + M_-^2 \, \Lambda_0}{M_-^2 \, \Lambda_+
+ M_+^2 \, \Lambda_0} \right)^{1/2}~, \nonumber\\
H &=& \frac{\Lambda_+ \, \Lambda_- - \Lambda_0^2}{\left( M_+^2 \,
\Lambda_- + M_-^2 \, \Lambda_0 \right)^{1/2} \, \left( M_-^2 \,
\Lambda_+ + M_+^2 \, \Lambda_0 \right)^{1/2} }~.
\labl{Hubble}
\end{eqnarray}
Even though both sectors have de Sitter backgrounds with
identical expansion rates, they are not identical,
since they are related by a ``physical'' (in the sense that it cannot
be removed by any coordinate reparameterization) conformal rescaling
with constant parameter $C\,$. We discuss the significance of this
rescaling in Section \ref{sc:matter}, where we introduce matter fields
coupled to the two gravity sectors.

A Minkowski background is obtained by a single tuning between 
the three ``cosmological constants'' $\Lambda_+ ,\, \Lambda_0 ,\,$ and
$\Lambda_- \,$ by requiring that the Hubble parameter \eqref{Hubble}
vanishes. The Minkowski solution is characterized by 
\begin{equation}
\Lambda_0 = \sqrt{\Lambda_+ \, \Lambda_-} \;\;,\qquad 
C = \left( \frac{\Lambda_-}{\Lambda_+} \right)^{1/4}~.
\labl{MinkFine}
\end{equation}
Notice that the parameter $C$ in this stationary background does not
depend on the values of $M_\pm\,$. Only for 
$\Lambda_+ = \Lambda_-\,$, the backgrounds become identical,
$C=1$. In the next section we investigate the spectrum of
perturbations around the Minkowski background characterized by an
arbitrary value of $C$.

\section{Perturbation spectrum of chiral gravity}
\labl{sc:perturs}

With the results of the Minkowski solution in mind, we investigate
the physical spectrum of the theory of chiral gravity in this
background. To this end we expand the vielbeins in terms of 
perturbations $\tilde F_\pm$ as 
\equ{
e_{\pm\, \gm}{}^a = C^{\pm 1/2} \Big( 
\gd^a_\gm + \tilde F_{\pm\, \gm\gn} \get^{\gn a}
\Big)~, 
\qquad 
\tilde F_{\pm\, \gm\gn} = F_{\pm\, \gm\gn} + B_{\pm\, \gm\gn}~. 
}
Metric perturbations are symmetric in the spacetime indices; however,
this is not required for the vielbein perturbations. Therefore, we
split their perturbations in symmetric $F_{\pm\, \gm\gn} = F_{\pm\,
\gn\gm}$ and anti--symmetric $B_{\pm\, \gm\gn} = -B_{\pm\, \gn\gm}$
ones. Using the diagonal Local Lorentz transformations we can require
that the anti--symmetric perturbations satisfy 
$B_{-\,\gm\gn} = - B_{+\, \gm\gn}\,$. Moreover, one can show that the
remaining anti--symmetric tensor perturbations are not dynamical: From
the linearized spin--connections 
\equ{
\go_{\pm\,}{}_{\gm}^{ab} = \get^{a\gs}\get^{b\gr} 
\Big( 
\der_\gm B_{\pm\, \gs\gr} 
- \der_\gs F_{\pm\, \gm\gr} 
+ \der_\gr F_{\pm\, \gm\gs} 
\Big)
}
we infer that the linearized curvatures 
\equ{
R_{\pm\,}{}^{ab}_{\gm\gn} = 
\Big( \get^{a\gs} \get^{b\gr} -  \get^{b\gs} \get^{a\gr}\Big) 
\Big( 
\der_\gm \der_{\gr} F_{\pm\, \gs\gn} 
- \der_\gn \der_{\gr} F_{\pm\, \gs\gm} 
\Big)~, 
}
are independent of the anti--symmetric parts $B_{\pm\, \gm\gn}$ of the
perturbations.~\footnote{We have confirmed that the anti--symmetric
tensor parts are non--dynamical by computing the quadratic
action of the theory as well. In this we differ from the conclusions
of~\cite{cham}, where equivalent kinetic terms were considered.} The
resulting equations of motion for the remaining anti--symmetric
contributions are trivial, and they imply that we can simply put
$B_{\pm\, \gm\gn} = 0\,$. Notice also that both the spin--connections
and the curvatures are independent of the conformal factor $C\,$.

To read off the spectrum of the theory we substitute these expansions
into the Einstein equations \eqref{EqMot} and obtain the set of
equations 
\equ{
\arry{rcl}{\dsp 
C\, M_+^2\, G^\ga_a (F_+) 
& = & \dsp 
\frac 1{24} 
\Big( 3 \gL_+ C^2 + 2 \gL_0 + 3 \gL_- C^{-2} \Big) \, 
\text{FP}^\ga_a(F_+ - F_-)~, 
\\[2ex] 
 C^{-1} \,M_-^2\, G^\ga_a (F_-) 
& = & \dsp 
\frac 1{24} 
\Big( 3 \gL_+ C^2 + 2 \gL_0 + 3 \gL_- C^{-2} \Big) \, 
\text{FP}^\ga_a(F_- - F_+)~.
}
\labl{PertEq}
}
Here the linearized Einstein tensor is given by 
\equ{
G_a^\ga (F) = R^\ga_a (F) - \frac 12 R(F) \, \gd^\ga_a~, 
}
where the explicit form of the linearized Ricci tensor is given in 
\eqref{linRicci}, and the Fierz--Pauli mass operator reads 
\equ{
\text{FP}^\ga_a(F) = 
2 \Big( 
\get^{\ga\gn} \, F_{a \gn} - \gd^\ga_a\, \get^{\gm\gn} F_{\gm\gn}
\Big)~. 
\labl{FPop}
}
The factor of two in this operator has been included for the
following reason: As observed in \eqref{EqCurvs}, the curvatures in
terms of the vielbein/spin--connection or the
metric/Christoffel--connection are equal; therefore, to identify the
graviton mass we should use the definition that corresponds to metric
perturbations 
\(
g_{\pm \, \gm\gn} = C^{\pm1}(\get_{\gm\gn} + h_{\pm\, \gm\gn})\,. 
\)
Using the fact that the vielbein perturbations are symmetric, we find the 
relation 
\equ{
h_{+\, \gm\gn} = 2 F_{+\, \gm\gn} 
+ F_{+\, \gm\ga} \get^{\ga\gb} F_{+\, \gb \gn}~, 
}
and similarly for the $-$ sector. Hence, to first order, \eqref{FPop}
is normalized precisely as a Fierz--Pauli mass term for the
perturbations of the metric.

To determine the mass eigenvalues we need to diagonalize the mass
terms in \eqref{PertEq}, while keeping the kinetic terms
diagonal. This is achieved by the transformation
\equ{
\pmtrx{ F_+ \\ F_-} = 
\frac 1{r + \frac 1r} 
\pmtrx{ 1 &  - \frac 1r \\ 1 & ~~r} 
\pmtrx{ F_0 \\ F_m}~, 
\qquad 
r = \frac {M_+}{M_-} C~. 
\label{diago}
} 
The perturbation equations for $F_0$ and $F_m$ decouple 
\equ{
G^\ga_a(F_0) = 0, 
\qquad 
G^\ga_a(F_m) = m_g^2 \, \text{FP}^\ga_a(F_m)~. 
}
Thus, $F_0$ describes a massless graviton, while $F_m$ a massive
graviton with the FP mass
\equ{
m_g^2 = \frac 13\,  {\gL_0}\, 
\frac{M_+^2 \,C + M_-^2 \, C^{-1}}{M_+^2\, M_-^2}~. 
\labl{GrMass}
}
We see that the graviton mass $m_g$ vanishes as $\gL_0 \rightarrow 0\,$, in which
case the the theory describes two massless gravitons (and a massless
vector and a massless scalar) as a consequence of the enlarged
covariance (as it is manifest in eq.~\eqref{acmetric}). The mass $m_g$
can be taken to be comparable with the present horizon scale, provided
$\Lambda_0$ is sufficiently small. It is worth noting that $m_g$
remains small even when one of the two masses $M_+$ or $M_-$ becomes
very large.

\section{Matter coupling} 
\labl{sc:matter}

We now address the natural question how the matter is coupled to the
two gravitons.  Our present discussion is far from complete, and it
is primarily aimed to determine the effective coupling strengths
(Planck masses) of the massless and massive gravitons identified in
the previous section by considering some specific examples. 
The simplest possibility is to consider two scalar fields $\gf_+$ and
$\gf_-$, which are only coupled to the $+$ and $-$ sector,
respectively. The scalar matter action reads concretely
\equ{
S_b = \int \d^4 x \left\{ \sqrt{-g_+} \Big[
- \frac 12 g_+^{\gm\gn} \der_\gm \gf_+ \der_\gn \gf_+ 
- \frac 12 m_+^2 \, \gf_+^2
\Big] 
+
\sqrt{-g_-} \Big[
- \frac 12 g_-^{\gm\gn} \der_\gm \gf_- \der_\gn \gf_- 
- \frac 12 m_-^2 \, \gf_-^2
\Big] 
\right\}~, 
}
where $m_+$ and $m_-$ are mass parameters. The scalars are not
canonically normalized in the static background~(\ref{MinkFine}). This
is easily accounted for by the wave function renormalization
\equ{
\tgf_\pm = C^{\pm 1/2} \, \gf_\pm~, 
\qquad 
\tm_\pm = C^{\pm 1/2} \, m_\pm~.
}
Notice, that the masses $m_\pm$ are also rescaled by the physical
conformal factor $C \,$. In this way, the energy momentum tensor is
identical to the standard one for Minkowski space
\equ{
\tT^{\gm\gn}_+ = 
\big( 
- \frac 12 \get^{\ga\gb} \der_\ga \tgf_+ \der_\gb \tgf_+
- \frac 12 \tm^2_+ \tgf_+^2
\big) \get^{\gm\gn} 
+ \get^{\gm\ga} \get^{\gn\gb}  \der_\ga \tgf_+ \der_\gb \tgf_+~,
}
and similarly for the energy--momentum tensor $\tT_-$ in the $-$ sector. 

For the moment we simply assume that our theory contains some form of
matter coupled to the $+$ and $-$ sectors encoded in the
energy--moment tensors $\tT_+$ and $\tT_-\,$, respectively. If we
assume that these forms of matter do not dominate, they are only a
source for the two gravitons via the linearized field
equations. The precise couplings follow from the
diagonalization~(\ref{diago}), and they read
\equ{
\arry{lcl}{\dsp 
G^\ga_a (F_0)
& =&  \dsp 
\frac 1{M_+ M_-} \, \tT_+{}^\ga_a  + \frac 1{M_+ M_-}\,  \tT_-{}^\ga_a~, 
\\[2ex] \dsp 
G^\ga_a (F_m)
- m_g^2 \, 
\text{FP}^\ga_a (F_m)
&= & \dsp 
- \frac 1{C M_+^2}\, \tT_+{}^\ga_a + \frac C{M_-^2}\, \tT_-{}^\ga_a~.
}
}
Notice that the massless graviton $F_0$ couples universally to the
canonically normalized $+$ and $-$ sectors of the theory. The coupling
of the massive graviton $F_m$ is not universal: In fact, because of
the opposite sign of the coupling of both sectors, the two sectors
feel a repelling force due to the massive graviton.

Hence, generally, the theory of chiral gravity contains both a
massless and a massive graviton in its spectrum. Each of them couples
to both the $+$ and the $-$ matter sectors, though only the massless
one has a universal coupling. It is interesting to consider the limit
in which either $M_+$ or $M_-$ becomes very large. In this limit the
massless graviton decouples (at the linear level) from matter, and the
massive graviton only couples to the $+$ sector if $M_-$ is taken
large, or to the $-$ sector if $M_+$ is large instead. Hence, assuming
that all the matter lives in the sector coupled to the massive
graviton, these limits describe a covariant nonlinear completion of
the FP mass term. Since only one linear coupling is nonvanishing, we
can use it do define an ``effective Planck mass'' for the (linearized)
gravitational interaction. More precisely, the graviton and Planck
masses are
\equ{
\arry{lll}{ \dsp 
m_g^2 \rightarrow \frac{\Lambda_0 \, C}{3 \, M_-^2}~, & \dsp 
M_P \rightarrow M_- \, C^{-1/2}~, & 
{\rm as } \;\; M_+ \rightarrow \infty~, \\[2ex] \dsp 
m_g^2 \rightarrow \frac{\Lambda_0}{3 \, C \, M_+^2}~, & \dsp 
M_P \rightarrow M_+ \, C^{1/2}~, &
{\rm as } \;\; M_- \rightarrow \infty~.
}
}
Notice that in either limit $m_g^2 \, M_P^2 \ra \gL_0 / 3 \,$. 

The introduction of fermionic matter requires more care. A detailed
study is beyond the aim of the present work, but some remarks are in
order. Fermions with positive chirality are coupled to the
spin--connection with positive chirality. For this reason, it is most
natural to construct their action only from the $+$ sector of gravity,
\equ{
S_f = \int \d^4 x\, e_+ \, \bgps_+ (e_+\inv)^\gm{}_a \gg^a 
\Big(\der_\gm + \go_{+\, \gm} \Big) \gps_+~, 
}
where $\go_{1+} = \go_{+\, \gm} \d x^\gm$ is given in equation
\eqref{eochi}. A similar action can be written down for the negative
chirality fermions, which couple to the $-$ sector of gravity. Like the
scalars considered above, these fermionic fields are not canonically
normalized in the Minkowski background described at the end of section
\ref{sc:background}. But, by a similar conformal rescaling, this
normalization is also obtained for the fermions, so that also their
energy--momentum tensors $\tT_\pm$ are constructed in a similar way to
the ones of the scalars.

Severe constraints have certainly to be expected, if the fermions of
the SM are embedded in this construction. If SM fermions
of different chiralities are coupled to the two different sectors,
this would presumably require taking $\Lambda_+ \simeq \Lambda_-$ and
$M_+ \simeq M_- \,$ with high precision. Alternatively, one can
construct the Standard Model only starting with one given chirality,
and then acting appropriately with the charge conjugation
operator. In this way, we can couple all the fermionic matter to
one given gravitational sector, and then proceed in analogy to the
scalar field case discussed above.

\section{Conclusions and outlook} 
\labl{sc:concl} 

The starting observation of this work is that the Fierz--Pauli
tensorial structure is immediately obtained by expanding the
determinant $e$ of the vielbein $e_\gm{}^a= \gd_\gm^a +
F_{\gm\gn}\get^{\gn a}$ to second order in
the perturbation. This is a quite unexpected result; it had been
observed since long that the same does not happen by expanding the
metric in its (first order) perturbation, but the simplest case of
the vielbein has (to our knowledge) never been noted so far. This
observation has led us to address two interesting and seemingly
unrelated questions: (i) how one can give a nonlinear completion of
the FP term which is covariant, and (ii) how one can construct a
theory of gravity that couples differently to positive and negative
chiralities.

Let us start from the second one. Since chirality is a property which
is naturally connected with local (tangent space) Lorentz 
transformations, we first reviewed the formulation of standard gravity
using the vielbein formalism, based on Clifford valued vielbein and
spin--connection one--forms. To obtain a chiral theory of gravity is
then straightforward: it requires two independent spin--connections
$\go_{\pm\,}{}_\gm^{ab}$ that are contracted with chiral Lorentz
generators $\gg_{ab}\frac {1\pm \gg_5}2\,$. To ensure that these
spin--connections are dynamically independent, two vielbeins
$e_{\pm\,_\gm}{}^a$ are needed to build to two Einstein--Hilbert terms
with Planck masses $M_\pm^2$. In principle there could have been two
more kinetic terms that describe the mixing of the spin--connection of
one sector with the vielbein of the other. We chose the structure of
the kinetic terms such that the standard expression of the
spin--connection  in terms of the vielbeins could be extended to both
sectors separately.

Enforcing reflection symmetry for both vielbeins separately, three
different ``cosmological constants'' could be introduced: in each
sector, the conventional one takes the form of a determinant,
$-\gL_\pm\, e_\pm = -\gL_\pm\, \sqrt{-\,g_\pm}\,$. The third one mixes
the two sectors, $\gL_0\, \langle e_+^2 e_-^2 \rangle\,$, and cannot
be represented as a determinant, see eq.~\eqref{newproduct}.  For
arbitrary (positive) values of these cosmological constants we found
de Sitter--like solutions in both sectors, with a common expansion
rate. However, the scale factors have different off--sets encoded by a
parameter $C \,$, which is function of the cosmological constants and
input Planck masses. A stationary background is obtained by a single
fine--tuning: $\gL_0^2 = \gL_+ \gL_-\,$. In this Minkowski background,
the parameter $C = (\gL_-/\gL_+)^{1/4}$ is not unity in general, hence
the conformal off--set remains between the two sectors.

The mix term does not break the covariance of the theory; there is a
massless graviton in the spectrum. The remaining degrees of freedom
combine in a non--dynamical anti--symmetric tensor, plus a massive 
graviton. As observed in the introduction, the FP mass term arises
naturally when expanding cosmological constant--like interaction in
vielbein perturbations. We have finally discussed how matter can be
included in this construction, and how it is coupled to the two
gravitons. Most relevant for our discussion is the fact that the
theory admits interesting limits, in which only the massive graviton
is coupled to matter at the linearized level. This construction
realizes a covariant nonlinear completion of the FP mass term. It had
been previously argued that a covariant formulation is not possible,
and that a background (Minkowski) metric has to be used. The
introduction of two gravitational sectors provides a way out to this
conclusion. An alternative possibility to obtain a covariant theory
is to use the Stuckelberg method ~\cite{ags}. As in bi--gravity
approaches, the spectrum of the theory is enlarged.

{Loosely speaking, the theory of chiral gravity can be viewed as a
bi--gravity theory since the set of gravitational fields is
doubled. The guideline we have followed, is to construct a model which
is as close as possible to the standard theory of gravity. For this
reason, we have only allowed for cosmological constant--like terms
that can be written down using the two vielbeins (the sets of all
possible terms has been further reduced by enforcing a parity symmetry
on the vielbeins). This approach can be compared with other
constructions were bi--gravity theories have been used to complete the
FP mass term; these models are typically characterized by a large
arbitrariness of the choice of potentials for the two gravitons, see
for instance ~\cite{kogan,isham,Padilla}. Although it is fair to say 
that also our criteria do not select a unique theory (and so, do not
allow for a predictive effective field theory, in the sense mentioned
in the Introduction), the model~(\ref{acmetric}) is an immediate and
simple generalization of the Einstein--Hilbert term of standard
gravity. Whether this may be of any practical advantage over other
choices requires to discuss the theory beyond the nonlinear level (we
hope to come back to this point in a separate publication).

The action~(\ref{acmetric}) appears to be particularly simple due to
the use of vielbeins rather than the two metrics. One could in
principle try to express the interaction $\langle e_+^2 \,e_-^2\rangle$ 
in terms of the two metrics. This would however lead to a rather
involved expression without any clear motivation. On the other hand,
the arguments we presented together with eq.~(\ref{expdet}) show that
a simple generalization of the cosmological constant term is unlikely
to have the FP limit at the linear level. This is a strong motivation
for the use of the vielbeins rather than the metric in the present
construction. 

The equivalence between chiral and the more often studied theories of 
bi-gravity (formulated using two metrics) is not trivial. 
The two vielbeins have also anti--symmetric components which,
as long as the two sectors are decoupled, can be removed by two
independent local Lorentz transformations. The mixing term $\langle
e_+^2 \, e_-^2 \rangle$ is invariant only under a combined Lorentz
transformation, therefore one of these anti--symmetric tensors cannot
be removed any longer. The perturbative calculations of
Section~\ref{sc:perturs} show that this field is not present in the
quadratic action for the perturbations. We expect that this is also
the case at the non--linear level. Indeed, a variant of the
St\"uckelberg formalism (along the lines of \cite{ags}) could be
employed to confirm that this anti--symmetric tensor is not dynamical
at any order. 

The potential presence of this additional field was already noted in
the work~\cite{cham}, which bears similarities with our
construction. Ref.~\cite{cham} considered a complex vielbein, which we
found to be equivalent to the chiral ones we have
introduced. In~\cite{cham}, it was claimed that the anti--symmetric
combination gives a dynamical field already at the quadratic
level. Unfortunately, the formulation given in~\cite{cham} is much
less tractable then the one we presented, since the kinetic terms for
the two different degrees of freedom do not appear to be
decoupled; for this reasons, the solutions in~\cite{cham} have
been given only perturbatively, and we believe that the claim that the
anti-symmetric field is physical (at the linear level) is
erroneous.~\footnote{In addition, the general relation between the FP
term and the expansion~(\ref{expdet2}) is not manifest (nor noted) 
in~\cite{cham}.}

There are several open issues left for future investigation. For
instance, we did not compute any experimental limit if different types
of matter are coupled to the two different gravitational sectors (see
however the discussion at the end of Section~\ref{sc:matter}).  We
also neglected the influence of matter on the evolution of the
background. Computing cosmological solutions in presence of matter
could help addressing the main motivation for massive gravity, namely
the present acceleration of the universe. In this respect, it is worth
noting that the massive graviton in our model can also mediate a
gravitational repulsion between different types of matter. However, from
what we already argued, it is clear that the
most relevant open questions are related to the nonlinear behavior of gravity in
this theory. We discussed in details the spectrum of linearized
perturbations around Minkowski background, showing how the massless
graviton decouples in certain limits. At the linear level, we have thus
only massive gravity of the FP form. However, the nonlinear structure
is now richer, and it could shed some light on open issues of massive gravity.
Computations of nontrivial backgrounds, for instance
the generalization of the standard black--hole solutions, may also
provide important information in this regard.}

\section*{Acknowledgments}

We would like to thank E.\ Dudas, A.\ Papazoglou, M.\ Pospelov, L.\ Sorbo, 
and A.~Vainshtein for very useful discussions. This work has been supported in
part by the Department of Energy under contract DE--FG02--94ER40823 at the
University of Minnesota.

\appendix 
\def\theequation{\thesection.\arabic{equation}} 
\setcounter{equation}{0}

\section{Technical details}
\labl{sc:details} 

We use the mostly plus convention for the metric, in particular the
Minkowski metric $\get_{ab}$ reads $\text{diag}(-1,1,1,1)$. We use 
indices $a,b,\ldots$ for the tangent space, and $\ga, \gb, \ldots$ 
to denote spacetime indices.  In particular, we use differentials
$\d x^\gm, \d x^\gn, \ldots$ that anti--commute 
\equ{
\d x^\gm \d x^\gn = - \d x^\gn \d x^\gm~, 
\qquad 
\d x^\ga \d x^\gb \d x^\gg \d x^\gd = \ge^{\ga\gb\gg\gd}\, \d^4 x~, 
}
where $\ge^{\ga\gb\gg\gd}$ is totally anti--symmetric, with 
$\ge^{0123} = 1\,$. 
In the definition of the forms we include appropriate symmetrization
factors. For example, for a two--form we write 
$B_2 = \frac 12 B_{\gm\gn} \d x^\gm \d x^\gn\,$.

The Clifford algebra is generated by $\gg_a$ that satisfies
\equ{
\{ \gg_a, \gg_b \} = 2\, \get_{ab}~, 
\qquad 
\gg_{ab} = \frac 12 [ \gg_a, \gg_b]~, 
\qquad 
\gg_5 = i\, \gg_0 \gg_1 \gg_2 \gg_3~. 
}
It follows that 
\equ{
\tr\, \gg_5 \gg_a \gg_b \gg_c \gg_d = 4 i \,\ge_{abcd}~, 
}
where $\ge_{abcd}$ is totally anti--symmetric with 
$\ge_{0123}=-1\,$.

The spin--connection is given in terms of the vielbein as 
\equ{
\arry{rcl}{
\go_\gr^{ab}(e) &= & - \frac 12 
\Big\{ 
\der_\gm e_\gn{}^c \, \get_{cd}\, e_\gr{}^d
+ \der^{}_{[\gm} e{}^c_{\gr]} \,\get_{cd}\, e_\gn{}^d
- (\gm \lra \gn) 
\Big\}\, \get^{am}\, \get^{bn}\, (e\inv)_m{}^\gm (e\inv)_n{}^\gn~, 
\\[2ex] 
R^{ab}_{\gm\gn}(\go) &=& 
\der^{}_{[\gm} \go_{\gn]}^{ab} 
+ \go^{ac}_{[\gm} \get_{cd} \go^{db}_{\gn]}~. 
}
\labl{solSpinCon} 
}
In the second line we have given the component form of the curvature 
defined in \eqref{SpinCurv}. The Christoffel--connection and curvature
read  
\equ{
\arry{rcl}{
\gG^\gl_{\gm\gn}(g) &=& \frac 12 g^{\gl\gr} 
\Big\{
- \der_\gr g_{\gm\gn} + \der_\gm g_{\gn\gr} + \der_\gn g_{\gm\gr}
\Big\}~, 
\\[2ex] 
R^\gl{}_{\gr\gm\gn}(\gG) &=& 
\der_\gm \gG^\gl_{\gn\gr} - \der_\gn \gG^\gl_{\gm\gr} 
- \gG^\gk_{\gm\gr} \gG^\gl_{\gn\gk} 
+\gG^\gk_{\gn\gr} \gG^\gl_{\gm\gk}~. 
}
\labl{ChrCon}
} 
Finally, the linearized Ricci tensor in the vielbein formalism is
expressed as 
\equ{
R^\ga_a(F) = \der_a \der^\gr (F_{\gr\gs} \get^{\gs\ga}) 
+ \der^\gr \der^\ga(F_{\gr a}) 
- \der_a \der^\ga (F_{\gr\gs} \get^{\gr\gs}) 
- \der_\gr \der^\gr (F_{a\gs} \get^{\gs\ga})~.
\labl{linRicci}
}


\begin{thebibliography}{99}


\bibitem{acc}
A.~G.~Riess {\it et al.}, Astron.\ J.\  {\bf 116}, 1009 (1998);
S.~Perlmutter {\it et al.}, Astrophys.\ J.\  {\bf 517}, 565 (1999);
D.~N.~Spergel {\it et al.}, Astrophys.\ J.\ Suppl.\  {\bf 148}, 175 (2003).

\bibitem{Arkani-Hamed:2003uy}
N.~Arkani-Hamed, H.~C.~Cheng, M.~A.~Luty and S.~Mukohyama,
JHEP {\bf 0405}, 074 (2004).

\bibitem{dps}
S.~L.~Dubovsky,
JCAP {\bf 0407}, 009 (2004); 
M.~Peloso and L.~Sorbo,
Phys.\ Lett.\ B {\bf 593}, 25 (2004).


\bibitem{fipa}
M.~Fierz and W.~Pauli,
Proc.\ Roy.\ Soc.\ Lond.\ A {\bf 173}, 211 (1939).

\bibitem{rudu}
V.~Rubakov, [hep-th/0407104];
S.~L.~Dubovsky,
JHEP {\bf 0410}, 076 (2004); 
B.~M.~Gripaios,
JHEP {\bf 0410}, 069 (2004). 

\bibitem{vdvz}
H.~van Dam and M.~J.~G.~Veltman,
Nucl.\ Phys.\ B {\bf 22}, 397 (1970);
V.~I.~Zakharov, JETP Lett. {\bf 12}, 312 (1970).

\bibitem{arkady1}
A.~I.~Vainshtein,
Phys.\ Lett.\ B {\bf 39}, 393 (1972).

\bibitem{arkady2}
C.~Deffayet, G.~R.~Dvali, G.~Gabadadze and A.~I.~Vainshtein,
Phys.\ Rev.\ D {\bf 65}, 044026 (2002).

\bibitem{ags}
N.~Arkani-Hamed, H.~Georgi and M.~D.~Schwartz,
Annals Phys.\  {\bf 305}, 96 (2003).

\bibitem{damour}
T.~Damour, I.~I.~Kogan and A.~Papazoglou,
Phys.\ Rev.\ D {\bf 67}, 064009 (2003).

\bibitem{bode}
D.~G.~Boulware and S.~Deser,
Phys.\ Rev.\ D {\bf 6}, 3368 (1972).

\bibitem{gagr}
G.~Gabadadze and A.~Gruzinov,
[hep-th/0312074].

\bibitem{hw}
A.~Hebecker and C.~Wetterich,
Phys.\ Lett.\ B {\bf 574}, 269 (2003); 
C.~Wetterich,
Phys.\ Rev.\ D {\bf 70}, 105004 (2004).

\bibitem{Duff}
M.~J.~Duff, C.~N.~Pope and K.~S.~Stelle,
Phys.\ Lett.\ B {\bf 223}, 386 (1989).

\bibitem{dgp}
G.~R.~Dvali, G.~Gabadadze and M.~Porrati,
Phys.\ Lett.\ B {\bf 485}, 208 (2000).

\bibitem{grs}
R.~Gregory, V.~A.~Rubakov and S.~M.~Sibiryakov,
Phys.\ Rev.\ Lett.\  {\bf 84}, 5928 (2000).

\bibitem{vanh}
J.~W.~van Holten,
``Supersymmetry And Supergravity: A Gauge Theory Approach. Part 1,''
WU B-85-3.

\bibitem{cham}
A.~H.~Chamseddine,
Phys.\ Rev.\ D {\bf 69}, 024015 (2004).

\bibitem{kogan}
T.~Damour and I.~I.~Kogan,
Phys.\ Rev.\ D {\bf 66}, 104024 (2002).

\bibitem{isham}
C.~J.~Isham, A.~Salam and J.~Strathdee,
Phys.\ Rev.\ D {\bf 3}, 867 (1971); 
A.~Salam and J.~Strathdee,
Phys.\ Rev.\ D {\bf 16}, 2668 (1977).

\bibitem{Padilla}
A.~Padilla,
Class.\ Quant.\ Grav.\  {\bf 21}, 2899 (2004). 


\end{thebibliography}
\end{document}